\documentclass[aps,pre,superscriptaddress,floatfix,showpacs,twocolumn]{revtex4}

\usepackage[dvips]{graphicx}
\usepackage{amssymb,amsmath}
\usepackage{bm}

\newcommand{\bridge}{\mathfrak{B}}

\newcommand{\mat}[1]{\mathsf{#1}}
\newcommand{\op}[1]{\mathcal{#1}}
\begin{document}

\title{Structure of molecular liquids: cavity and bridge functions of the hard 
spheroid fluid}

\author{David L. Cheung}
\email{david.cheung@warwick.ac.uk}
\affiliation{Department of Physics and Centre for Scientific Computing \\
  University of Warwick, Coventry, CV4 7AL, UK}

\author{Lucian Anton}
\email{lucian.anton@manchester.ac.uk}
\affiliation{School of Chemical Engineering and Analytical Science \\
  University of Manchester, Sackville Street, Manchester,  M60 1QD,
  UK}

\author{Michael P. Allen}
\email{m.p.allen@warwick.ac.uk}
\affiliation{Department of Physics and Centre for Scientific Computing \\
  University of Warwick, Coventry, CV4 7AL, UK}

\author{Andrew J. Masters}
\email{andrew.masters@manchester.ac.uk}
\affiliation{School of Chemical Engineering and Analytical Science \\
  University of Manchester, Sackville Street, Manchester,
  M60 1QD,
  UK}

\begin{abstract}
  
  We present methodologies for calculating the direct correlation
  function, $c(1,2)$, the cavity function, $y(1,2)$, and the bridge
  function, $b(1,2)$, for molecular liquids, from Monte Carlo
  simulations. As an example we present results for the isotropic hard
  spheroid fluid with elongation $e=3$. The simulation data are
  compared with the results from integral equation theory.  In
  particular, we solve the Percus-Yevick and Hypernetted Chain
  equations. In addition, we calculate the first two terms in the
  virial expansion of the bridge function and incorporate this into
  the closure. At low densities, the bridge functions calculated by
  theory and from simulation are in good agreement, lending support to
  the correctness of our numerical procedures. At higher densities,
  the hypernetted chain results are brought into closer agreement with
  simulation by incorporating the approximate bridge function, but
  significant discrepancies remain.

\end{abstract}
\pacs{61.20.Gy, 61.20.Ja, 05.20.Jj}

\maketitle

\section{\label{sec:intro}Introduction}

The equilibrium properties of homogeneous fluids of spherical
particles have been extensively studied both by theory and simulation
and a great deal is now known about the thermodynamic properties and
the fluid structure \cite{hansenmcdonald,graygubbins}. Simulation has
been used to calculate the total and direct correlation functions
\cite{groot1987a}, the cavity function \cite{patey1977a,ballance1985a}
and the bridge function
\cite{chapman1992a,chapman1994a,malijevsky2002a}. On the theoretical
side, integral equation theory (IET) is now capable of making some
very accurate predictions. Percus-Yevick (PY) and Hypernetted Chain
(HNC) theories have now been extended, for example, by mixing closures
so as to obtain identical virial and compressibility equations of
state \cite{rogers1984a,hansen1986a}. An alternative approach has been
to incorporate approximate forms for the bridge function in the HNC
closure. These may take the form of a low-order virial expansion, a
bridge function from a reference fluid or an approximate closure
relation \cite{rosenfeld1979a,patey1989a,rast1999a}.  While there is
still work to be done, especially perhaps on a fundamental treatment
of the bridge function, the foundations are rather solid. As a
consequence, one is in a good position to construct good density
functionals to describe inhomogeneous fluids. A key ingredient of a
density functional is an assumed form for the inhomogeneous direct
correlation function. It is clearly reassuring if this quantity
reduces to the known homogeneous function in the uniform limit and,
for spherically symmetric particles, this is a test one may apply.
 
The equilibrium properties of isotropic fluids of non-spherical
particles are less well-characterised. There are relatively few
simulation studies on the direct correlation function
\cite{allen1995a} and little data for the cavity or bridge functions
have been published (site-site functions have been calculated for hard
sphere dimers and water \cite{lue95,lee2001a,lee2001b} while the first
bridge diagram has been calculated for the hard spherocylinder fluid
for a number of fixed orientations \cite{malijevsky1994a}). The PY and
HNC equations have been solved for axially symmetric particles (e.g.\
hard ellipsoids, hard spherocylinders and truncated hard spheres) and
the general conclusion is that HNC is superior to PY for significantly
aspherical particles, but that there is still a substantial
discrepancy between theory and simulation, especially at high density
\cite{patey1985a,patey1987a,patey1988a,latz1999a}. There have been
some attempts to go beyond HNC. \citet{pospisil1993a} have
investigated the use of a modified Verlet-bridge closure and have
reported improved results.  \citet{singh96} applied a non-spherical
version of the Rogers-Young method of mixing PY and HNC closures,
again obtaining results for spheroids in good agreement with
simulation.  Nevertheless the number of such studies is relatively
small and, as yet, we do not have sufficient simulation and
theoretical studies to claim a foundation to rival that enjoyed by
spherical particles.

In this paper we try to address some of these issues, using Monte
Carlo (MC) simulations and IET. On the simulation front, we present
methodologies for calculating the direct correlation, cavity and
bridge functions for isotropic fluids of axially symmetric particles
using advanced MC techniques. These methods are used to calculate the
molecular correlation functions for a fluid of hard spheroids with
major axis of length $a$ and minor axis of length $b$. We focus here
on an elongation $e=a/b=3$, and present results for a range of
densities in the isotropic phase. IET is adapted for fluids of
anisotropic particles using invariant expansions of the correlation
functions \cite{blum1972a,blum1972b} and efficient numerical
algorithms
\cite{patey1987a,patey1988a,talbot90,labik91,ram94,kinoshita91}. In
particular we use the relaxation method of \citet{ng74} to provide a
robust and easily-programmable algorithm for numerically solving the
integral equations. We also examine some analytical properties of the
cavity function for non-spherical hard particles and calculate the
first two terms in the virial expansion of the bridge function.

The paper is organised as follows. In section \ref{sec:gen_form}, we
give the basic equations relating the correlation functions studied in
this article. Section \ref{sec:sim} describes the simulation methods
used for the calculation of the cavity function and bridge
function. In Section \ref{sec:iet} we present some technical details
of the numerical solution of the IET using the method of \citet{ng74}
and the Monte Carlo procedure used to compute the bridge diagrams. In
Section \ref{sec:res} the results of simulations and IET are compared
and discussed. The general conclusion of our study and some future
avenues of work are given in Section \ref{sec:conc}. Some more
technical details of the Monte Carlo algorithm for the calculation of
the cavity function are presented in the Appendix.

\section{\label{sec:gen_form} General formalism}

The structure of a fluid may be described at a two-particle level by
the total correlation function (TCF) $h(1,2)=g(1,2)-1$ (where $g(1,2)$
is the pair distribution function) or the direct correlation function
(DCF) $c(1,2)$. These are linked via the Ornstein-Zernike (OZ)
equation, which for a homogeneous fluid of axially symmetric molecules
is \cite{graygubbins,hansenmcdonald}
\begin{equation}\label{eqn:oz}
  h(1,2)=c(1,2)+\frac{\rho}{4\pi}\int\;d3\;c(1,3)h(3,2)
\end{equation}
where $\rho$ is the number density and, as is traditional,
$(i)\rightarrow(\bm{r}_i, \bm{u}_i)$. Here $\bm{r}_i$ denotes the
centre of mass position of particle $i$ whilst $\bm{u}_i$ represents a
unit vector along the particle's symmetry axis.
 
To determine $h(1,2)$ and $c(1,2)$, Eq.~\eqref{eqn:oz} is usually
supplemented by an approximate closure relation. These take the form
\begin{align}
  c(1,2)&=(1+h(1,2))\bigl(1-\exp[\beta V(1,2)]\bigr)
  &&\text{PY}\label{eqn:oz_py1} \\
  c(1,2)&=h(1,2)-\log\bigl[1+h(1,2)\bigr]-\beta V(1,2)
  &&\text{HNC} \label{eqn:oz_hnc1}
\end{align} 
where $V(1,2)$ is the intermolecular potential and
$\beta=1/k_\text{B}T$. 

The exact closure relation can be written as follows
\cite{hansenmcdonald}
\begin{equation}\label{eqn:oz_close}
  y(1,2)=\exp\left\{h(1,2)-c(1,2)+b(1,2)\right\}
\end{equation}
where $b(1,2)$ is the bridge function and $y(1,2)$ is the cavity or
background correlation function defined by the relation: 
\begin{equation}\label{eqn:y_def}
y(1,2)=g(1,2)\exp\left[\beta
V(1,2)\right] \ .
\end{equation}
Eq.~\eqref{eqn:oz_close} may be regarded as a definition of $b(1,2)$
and the approximate closure relations may be regarded as
approximations to the unknown $b(1,2)$. In particular, the PY and HNC
closures, Eqs.~\eqref{eqn:oz_py1},\eqref{eqn:oz_hnc1} respectively,
correspond to
\begin{subequations}
\label{eqn:b_closures}
\begin{align}
  b(1,2) &=\eta(1,2)-\log(1+\eta(1,2)) &&\text{PY}
\label{eqn:py_bridge}
\\ b(1,2) &=0 &&\text{HNC}
\label{eqn:oz_hnc}
\end{align}
where $\eta(1,2)=h(1,2)-c(1,2)$.  

The bridge function may be expressed as a virial expansion
\begin{equation*} %\label{eqn:vir_bridge}
  b(1,2)=\sum_{n\ge2}\rho^n \bridge_n(1,2) \ ,
\end{equation*}
where $\bridge_n(1,2)$ are the bridge diagrams. In principle, this
provides a route for the exact calculation of the bridge function,
but in practice it is only feasible to calculate low-order terms,
as has been done for hard spheres \cite{patey1989a,rast1999a}.
In this paper we use the two lowest-order estimates of the bridge function
for hard spheroids
\begin{align}
  b_2(1,2) &= \rho^2 \bridge_2(1,2) && \text{HNC+B2} 
\label{eqn:vir_bridge2}
\\
  b_3(1,2) & = \rho^2 \bridge_2(1,2) + \rho^3 \bridge_3(1,2) &&\text{HNC+B3}
\label{eqn:vir_bridge3}
\end{align}
\end{subequations}
to extend the HNC closure relation. In this paper we investigate all four
closure relations, i.e.\  Eq.~\eqref{eqn:oz_close} with the bridge function
specified by one of Eqs.~\eqref{eqn:py_bridge}--\eqref{eqn:vir_bridge3}.

The numerical solution of the integral equation and MC calculation are
based upon the expansion of two-particle functions in a basis set of
rotational invariants \cite{blum1972a,graygubbins}:
\begin{align}
  F(1,2)&=\sum_{mn\ell} F^{mn\ell}(r)\Phi^{mn\ell}(\bm{u}_1,\bm{u}_2, \bm{u}_r)
\ ,\label{eqn:labf} \\ \Phi^{mn\ell}(\bm{u}_1,\bm{u}_2, \bm{u}_r)&= 4\pi \sum_{\chi_1 \chi_2
\chi_r} \left(
\begin{array}{ccc} m & n & \ell  \\ \chi_1 & \chi_2 & \chi_r \end{array} \right) \notag\\ 
& \times Y_{m\chi_1}(\bm{u}_1)Y_{n\chi_2}(\bm{u}_2) C_{\ell\chi_r}(\bm{u}_r) \ ,
\label{eqn:rotinv}
\end{align}
where $r$ is the intermolecular distance; $\bm{u}_r$ is a unit vector
along the intermolecular vector, $\bm{u}_1$, $\bm{u}_2$ are the
orientations of the molecules in a given system of coordinates
(`laboratory frame'), $Y_{m\chi}(\bm{u})$ are the spherical harmonics
functions, $C_{m\chi}(\bm{u})=(4\pi/(2m+1))^{1/2}Y_{m\chi}(\bm{u})$ and
\begin{equation*}
\left(
\begin{array}{ccc} m & n & \ell \\ \chi_1 & \chi_2 & \chi_r \end{array} \right)
\end{equation*}
are the standard $3j$ symbols.

Some quantities of interest are easier to compute in a system of
coordinates whose $z$-axis lies along the intermolecular vector
(`molecular frame').  The expansion in the molecular frame has the
form:
\begin{align}
  F(1,2)&=4\pi \sum_{mn\chi}
  F_{mn\chi}(r)Y_{m\chi}(\tilde{\bm{u}}_1)Y_{n\bar\chi}(\tilde{\bm{u}}_2) \ ,
\end{align}
where $\bar\chi=-\chi$.
The two sets of coefficients are connected through the
$\chi$-transform and its inverse:
\begin{subequations}
\label{eqn:chi_and_inv}
\begin{align}
\label{eqn:chi}
  F_{mn\chi}(r)& =\sum_{\ell}\left(\begin{array}{ccc} m & n & \ell \\ \chi &
  \bar\chi & 0 \end{array}\right) F^{mn\ell}(r)\\
\label{eqn:invchi}
 F^{mn\ell}(r) & = (2\ell+1)\sum_{\chi}\left(\begin{array}{ccc} m & n & \ell \\ \chi &
  \bar\chi & 0 \end{array}\right) F_{mn\chi}(r)\ .
\end{align}
\end{subequations}

\section{\label{sec:sim}Simulation method}

\subsection{\label{ssec:sim_dcf}Direct correlation function}

The total correlation function may be determined directly from
simulation through the pair distribution function $g(1,2)$. The
spherical harmonic coefficients are determined as usual from
\cite{tildesley1976a}
\begin{equation}\label{eqn:g_mnx}
  g_{mn\chi}(r)=4\pi g_{000}(r)\left<Y^*_{m\chi}(\bm{u}_1)
    Y^*_{n\bar{\chi}}(\bm{u}_2)
    \right>_r \ ,
\end{equation}
where $Y_{m\chi}(\bm{u})$ is a spherical harmonic, $\bar{\chi}=-\chi$,
$g_{000}(r)$ is the pair distribution function of the particle centres, and
the angled brackets denote an average over all molecules in the shell
$[r,r+\delta r]$. These coefficients are defined in the molecular frame
described in Section~\ref{sec:gen_form} \cite{graygubbins}. From
Eq.~\eqref{eqn:g_mnx} it follows that
$h_{mn\chi}(r)=g_{mn\chi}(r)-\delta_{m0}\delta_{n0}\delta_{\chi0}$.

The direct correlation function may be found from the measured total
correlation function in two ways. In reciprocal space, using the
molecular frame expansion, the Ornstein-Zernike equation becomes
\begin{equation}
  \tilde{h}_{mn\chi}(k)=\tilde{c}_{mn\chi}(k)+(-1)^\chi\rho
  \sum_{j}\tilde{h}_{mj\chi}(k)\tilde{c}_{jn\chi}(k)
\end{equation}
where $\tilde{f}(k)$ is the (three-dimensional) Fourier transform of a
function $f(r)$. The structure of this equation with respect to the
first two indices leads to a matrix notation
\begin{equation}
  \tilde{\mat{H}}_\chi(k)=\tilde{\mat{C}}_\chi(k)+
  (-1)^\chi\rho\tilde{\mat{H}}_\chi(k)
               \tilde{\mat{C}}_\chi(k) \ .
\end{equation}

$c(1,2)$  may also be obtained via a real-space factorization
 \cite{graygubbins,blum1981a,blum1990a}. It is possible to write
\begin{subequations}
\label{eqn:oz_real}
\begin{align}
  r\hat{\mat{C}}_\chi(r) &=-\mat{Q}_\chi'(r)+2\pi(-1)^\chi\rho
  \int_r^R\;ds\mat{Q}_\chi'(s)\mat{Q}_\chi^\text{T}(s-r)
\label{eqn:oz_real1}
\\
  r\hat{\mat{H}}_\chi(r) &=-\mat{Q}_\chi'(r)+2\pi(-1)^\chi\rho
  \int_0^R\;ds(r-s)\hat{\mat{H}}_\chi'(r-s)\mat{Q}_\chi(s)
\label{eqn:oz_real2}
\end{align}
\end{subequations}
where the new matrix $\mat{Q}_\chi(r)$ has been introduced,
$\mat{Q}_\chi'(r)=d\mat{Q}_\chi(r)/dr$, and $\mat{Q}_\chi^\text{T}(r)$
is the transpose of $\mat{Q}_\chi(r)$. 

 The so-called `hat' transform giving the functions $\hat{\mat{H}_\chi}$,
$\hat{\mat{C}_\chi}$, that appear in Eqs.~\eqref{eqn:oz_real} is defined in
the laboratory frame
\begin{equation}\label{eqn:hat_trans}
  \hat{f}^{mn\ell}(r)=f^{mn\ell}(r)-\int_r^\infty ds\;
  s^{-1}f^{mn\ell}(s)P_\ell^\text{e}(r/s)
\end{equation}
where $P_\ell^\text{e}(x)=x^{-1}dP_\ell(x)/dx$ and $P_\ell(x)$ are
Legendre polynomials. A $\chi$-transform, Eq.~\eqref{eqn:chi}, then
converts the functions to the required molecular frame.  It is assumed
that a separation $R$ exists such that $\mat{Q}_\chi(r)=0$ and
$\mat{C}_\chi(r)=0$ for all $r>R$. Equation \eqref{eqn:oz_real2} is
solved iteratively to find $\mat{Q}_\chi(r)$ from the functions
$\mat{H}_\chi(r)$ determined in the simulation. Once this procedure
has converged, Eq.~\eqref{eqn:oz_real1} is used to determine
$\hat{\mat{C}}_\chi(r)$. At very small $r$ this involves the
difference between two large quantities possibly leading to numerical
difficulties. These may be avoided by a procedure outlined previously
\cite{allen1995a}, finding $\hat{\mat{C}}_\chi(0)$ from
\begin{align}
  \hat{\mat{C}}_\chi(0)-\hat{\mat{H}}_\chi(0)&=(-1)^\chi
  2\pi\rho
  \left\{\left(\int_0^Rdr\;r\hat{\mat{H}}_\chi(r)\mat{Q}_\chi'(r)
  \right .\right . \notag \\ & -
  \mat{Q}_\chi'(r){\mat{Q}_\chi^\text{T}}'(r)\Big)-
  \mat{Q}_\chi(0)\mat{Q}_\chi^\text{T}(0)\Bigg\}
\end{align}
and the $\hat{\mat{C}}_\chi(r)$ for $r\rightarrow0$ are then
determined by interpolation.

\subsection{\label{ssec:sim_cav}Cavity correlation function}

There are two methods for the calculation of the cavity function,
Eq.~\eqref{eqn:y_def}, either by a direct simulation of two
non-interacting cavity particles \cite{patey1977a} or through the 
test-particle method based on Henderson's equation
\cite{henderson1983a}. The first of these methods is more useful for
large cavity separations while the second is better as
$r\rightarrow0$.

\subsubsection{\label{sssec:sim_cav_dir}Direct simulation method}

The direct simulation method follows from the observation that for the
hard particle fluid, the cavity function may be identified as the pair
distribution function for a pair of non-interacting cavities
\cite{meeron1968a}. In a MC simulation it is convenient to constrain
the two cavities to be within a given range of separations
$r_{12}$. Even so, in a normal MC simulation the probability
distribution $P_\text{cav}(1,2)$ is likely to vary rapidly with
separation $r_{12}$, leading to poor sampling in the regions where the
function is relatively small.  To circumvent this problem, the
umbrella sampling technique is employed \cite{frenkelsmit}. The
$r$-separation of the cavities is divided into a set of overlapping
windows. Within each window a weight function $w(r_{12})$ is
introduced into the Monte Carlo moves; this function is iteratively
refined so as to produce a flat sampled probability distribution. This
weight may be subsequently removed to give the true probability
distribution for each window and the full distribution is recovered
using the self-consistent histogram method \cite{ferrenberg1989a}.

The cavity function is, to within a multiplicative constant, equal to
$P_\text{cav}(1,2)/r^2_{12}$. When the cavity particles are
constrained this constant cannot be determined directly
\cite{chapman1992a}. However it may be found by enforcing the
condition that $y(1,2)=1+h(1,2)$ when outside the overlap region
\cite{patey1977a}.

For this scheme to be effective a good choice of the weighting
function is needed.  For hard spheres a good choice proved to be an
analytic approximation to $y(r)$ \cite{patey1977a}. Here we employ a
more general method based on the Wang-Landau method
\cite{landau2001a,debenedetti2002a}. Briefly, this is an iterative
method that updates an initial guess to the weight function using a
decreasing modification factor. Full details are given in the
appendix. The implementation used here is similar in spirit to the
extended density of states method (EDOS)
\cite{depablo2002a,depablo2004a}. The spherical harmonics coefficients
$y_{mn\chi}(r)$ are found in the same way as those for the pair
distribution function $g_{mn\chi}(r)$.

\subsubsection{\label{sssec:sim_cav_test}Test particle method}

In the canonical ensemble, Henderson's equation for a system containing $N$
molecules may be written \cite{chapman1992a}
\begin{align}
  y(0,1)&=\exp\left[\beta\mu_\text{ex}\right] \notag\\
  &\times\left\langle\exp\!\left(-\sum_{j\ge 2}^N\beta(
  V(0,j)\right)\right\rangle_{N,V,T}, \label{eqn:hend1}
\end{align}
where $\mu_\text{ex}$ is the excess chemical potential and the angled
brackets denote an ensemble average over particles $1,\dots, N$.  The
term in the angled brackets corresponds to the Boltzmann factor of a
molecule 0 with the interaction with another molecule 1
neglected. This may intuitively be equated to a fluid consisting of 2
non-interacting cavities and $N-1$ other molecules. Additionally this
is also equivalent to the calculation of the acceptance criterion in a
Metropolis MC simulation - the interaction between a hypothetical
molecule with position $\bm{r}_0$ and orientation $\bm{u}_0$ and every
molecule in the system apart from $1$ is the quantity that is
calculated when an attempt is made to move molecule $1$ to position
$\bm{r}_0$ and orientation $\bm{u}_0$. So as any molecule in the
system may be labelled $1$ the quantity in the angled brackets in
Eq.~\eqref{eqn:hend1} is calculated for every attempted MC move. This
fact has been used in previous studies of atomic fluids
\cite{chapman1992a,chapman1994a} allowing the calculation of $y(0,1)$
at essentially no extra cost. However, for molecular fluids, where the
maximum angular displacement in a Monte Carlo simulation may be much
smaller than $2\pi$, this would lead to poor sampling of the angular
dependence $y(0,1)$. Hence the calculation of $y(0,1)$ proceeds by
inserting a number of test particles (labelled 0) in the vicinity of
each molecule in the simulation (labelled 1).  The Boltzmann factor
(neglecting the interaction between 0 and 1) is then calculated. For
hard molecules, as in the present case, this is simply 0 if the test
particle overlaps with any other molecule (excluding molecule 1) or 1
if there are no overlaps.

The spherical harmonic coefficients $y_{mn\chi}(r)$ are given by
\begin{align}
  y_{mn\chi}(r) &=
  (4\pi)^{-1}\int\;d\bm{u}_0d\bm{u}_1\;y(0,1)Y_{m\chi}^*(\bm{u}_0)
  Y^*_{n\bar{\chi}}(\bm{u}_1) \notag\\ &=
  (4\pi)^{-1}\exp\left[\beta\mu_\text{ex}\right]
  \left\langle\exp\left(-\sum_{j\ge 2}^N\beta V(0,j)\right)
  \right. \notag \\ & \times
  Y_{m\chi}^*(\bm{u}_0)Y^*_{n\bar{\chi}}(\bm{u}_1) \Bigg\rangle_r \ ,
  \label{eq:ymnx}
\end{align}
where the angled brackets denote averages over test particle
insertions in the range $[r,r+\delta r]$.

\subsection{\label{ssec:sim_bridge}Bridge function}

Once spherical harmonic expansions for $h(1,2)$, $c(1,2)$ and $y(1,2)$
have been determined, the final step is to invert
Eq.~\eqref{eqn:oz_close} for $b(1,2)$. While the presence of the
exponential on the right hand side of Eq.~\eqref{eqn:oz_close} is
troublesome for the spherical harmonics expansions, it may be easily
circumvented \cite{patey1985a}.  Taking the logarithm and
differentiating Eq.~\eqref{eqn:oz_close} with respect to $r$ gives
\begin{equation}\label{eqn:bridge1}
  \frac{\partial y(1,2)}{\partial r} =
  y(1,2)\left[
    \frac{\partial h(1,2)}{\partial r} - 
    \frac{\partial c(1,2)}{\partial r} +
    \frac{\partial b(1,2)}{\partial r}
    \right].
\end{equation}
Inserting the spherical harmonic expansions of the pair functions
 and integrating over angles gives\cite{graygubbins}
\begin{align}
  \frac{d y_{mn\chi}(r)}{d r} &= 4\pi\sum_{\substack{m' n' \chi' \\ m''n'' \chi''}}
  \Gamma^{mm'm''}_{\chi\chi'\chi''}
  \Gamma^{nn'n''}_{\bar{\chi} \bar{\chi}'\bar{\chi}''}y_{m''n''\chi''}(r) 
  \nonumber\\
  &\times
  \frac{d}{d r}\left[h_{m'n'\chi'}(r)-c_{m'n'\chi'}(r)+b_{m'n'\chi'}(r)\right] 
  \label{eqn:bridge2}
\end{align}
where 
\begin{align}
\Gamma^{mm'm''}_{\chi\chi'\chi''} &=\int d\bm{u} \;
  Y^*_{m\chi}(\bm{u})Y_{m'\chi'}(\bm{u})Y_{m''\chi''}(\bm{u})
  \nonumber\\ &=\sqrt{\frac{(2m'+1)(2m''+1)}{4\pi(2m+1)}}
  C(m'',m',m;0,0,0)\nonumber \\ & \times C(m'',m',m;\chi'',\chi',\chi) \ ,
  \label{eqn:bridge3}
\end{align}
and where $C(m'',m',m;\chi'',\chi'\chi)$ are Clebsch-Gordan coefficients.
Eq.~\eqref{eqn:bridge2} can be solved using standard numerical methods
\cite{numericalrecipies} for the derivatives $d b_{mn\chi}(r)/d r$, and these are
integrated numerically to give the bridge function components $b_{mn\chi}(r)$.

\subsection{Simulated system}

The simulated system consists of a fluid of hard prolate spheroids of
elongation $e=a/b=3$. This is a common model for molecular fluids and
liquid crystals and along with similar models such as hard
spherocylinders has been well studied \cite{allen1993a}.

For the calculation of $h(1,2)$, systems of $2048$ molecules were simulated
using constant $NVT$ MC simulations. Data for the calculation of $h(1,2)$ were
gathered every $500$ MC sweeps (each sweep is on average $1$ attempted
translation and $1$ attempted rotation per molecule) over a total of $5\times
10^5$ MC sweeps. The $c_{mn\chi}(r)$ coefficients were then calculated from the
$h_{mn\chi}(r)$ coefficients following Sec.~\ref{ssec:sim_dcf}. The
spherical harmonics expansions for the pair functions were truncated
at $m_\text{max}=n_\text{max}=8$ and the grid spacing $\delta r =0.01b$.

For the calculation of $y(1,2)$, systems of $512$ molecules, including
$2$ cavity molecules, were simulated (smaller systems are sufficient
for the calculation of $y(1,2)$ as its long-range behaviour is
identical to $h(1,2)$).  The $r$ separation between the cavity
particles was split into overlapping windows covering
$r/b=[0.03,0.50]$, $[0.20,1.20]$, $[1.00,2.00]$, $[1.80,2.80]$, and
$[2.60,3.60]$. In each window the weight function was determined over
at least $15$ iterations (see appendix for details). Once the final
weight function was determined, $y(1,2)$ data were gathered over a
total of $2\times 10^7$ MC sweeps.  Error estimates were made by
splitting this into 4 subruns.  $y(1,2)$ was calculated for the region
$r/b=[0,0.15]$ using the test particle insertion method
(Sec.~\ref{sssec:sim_cav_test}). $h(1,2)$ and $y(1,2)$ have been
calculated at reduced densities
$\rho^*=\rho/\rho_\text{cp}=0.10,0.20,0.30,0.40,0.50$ where
$\rho_\text{cp}=\sqrt{2}/(ab^2)$ is the close-packed density.

\section{\label{sec:iet}Integral Equation Theory}

To solve the integral equations in the isotropic phase we have used
the standard rotationally invariant decomposition of the angular part
of the correlation functions as discussed in detail in
Refs.~\cite{patey1987a,talbot90}. The solution is calculated
iteratively with the help of the method of \citet{ng74} that yields
fast convergence even at densities close to those where no real
solution exists.

We describe in short the \citeauthor{ng74} method as applied to the
hard spheroid fluid.  An iteration step in the \citeauthor{ng74}
method is done using as input a linear combination of the $p$
functions obtained in the $p$ previous steps. The coefficients of the
linear combination are calculated from a smallest displacement
condition.

An iteration has the generic form
\begin{align}
  f_{i+1}(1,2)&=\op{O}[t_i(1,2)] 
  \label{eq:nginit} \\
\text{with}\quad t_i(1,2)&=f_i(1,2)-\sum_{m=1}^{p} \alpha_{i,m} \Delta
f_{i,m}(1,2)
\end{align}
where $\op{O}[f]$ is the iteration operator, $f_i(1,2)$ is the $i^\text{th}$
 iteration result and $\Delta f_{i,m}(1,2)=f_i(1,2)-f_{i-m}(1,2)$. At each
 iteration step the scalars $\alpha_{i,m}$ are computed from the minimum
 condition of the following functional:
\begin{equation}
 \int d2 \bigl[f_{i+1}(1,2)-t_i(1,2)\bigr]^2 \; . \label{eq:ngmin}
\end{equation}

Close to the solution we assume that the differences $\Delta
f_{i,m}(1,2)$ are small and we expand Eq.~\eqref{eq:nginit} up to the
first order:
\begin{multline}
  \op{O}\bigl[f_i(1,2)-\sum_{m=1}^{p} \alpha_{i,m} \Delta f_{i,m}(1,2)\bigr] \\  \approx
  \op{O}[f_i(1,2)]-\sum _{m=1}^{p} \alpha_{i,m} \frac{\partial
  \op{O}[f_i(1,2)]}{\partial f_i(1,2)}\Delta f_{i,m}(1,2) \; . \label{eqn:O_ex}
\end{multline}
 The coefficients $\alpha_{i,m}$ that satisfy the approximated minimum
 condition, Eq.~\eqref{eq:ngmin}, are the solutions of a linear system of
 equations
\begin{equation}
  \sum_{m=1}^p a_{km}\alpha_{i,m}=b_k \;, \quad k=1 \ldots p \; ,
\end{equation}
where the coefficients $a_{km}$ and $b_k$ are determined from the following
equations:
\begin{align}
  a_{km} & =\int d2 \; \delta \op{O}[f_i(1,2)]_k \delta
  \op{O}[f_i(1,2)]_m \\ b_k & = \int d2\;
  \bigl(\op{O}[f_i(1,2)]-f_i(1,2)\bigr) \delta \op{O}[f_i(1,2)]_k \; ,
\end{align}
and
\begin{equation*}
 \delta \op{O}[f_i(1,2)]_k=\frac{ \partial \op{O}[f_i(1,2)]}{\partial
f_i(1,2)}\Delta f_{i,k}(1,2) \; .
\end{equation*}

In our case the nonlinear operator $\op{O}[\cdot]$ has the form
\cite{patey1987a}
\begin{equation}
 \op{O}[\eta]= \lambda(1,2)( -1-\eta(1,2))+(1-
  \lambda(1,2))c_{\text{cl}}(1,2) \label{eq:op_her} \ ,
\end{equation}
where $\lambda(1,2)$ has the value $1$ if spheroids $1$ and $2$ overlap
and the value $0$ if they do not. $c_{\text{cl}}(1,2)$ is given by 
\begin{subequations}
\label{eq:closure}
\begin{align}
%  c_{\text{cl}}(1,2) 
 & 0  && \text{PY} 
\label{eqn:closure_py}\\
  & \exp\bigl(\eta(1,2)\bigr)-\eta(1,2)-1 && \text{HNC}
\label{eqn:closure_hnc} \\
  & \exp\bigl(\eta(1,2)+b_2(1,2)\bigr)-\eta(1,2)-1  && \text{HNC+B2} 
\label{eqn:closure_vir2} \\
  & \exp\bigl(\eta(1,2)+b_3(1,2)\bigr)-\eta(1,2)-1  && \text{HNC+B3} 
\label{eqn:closure_vir3}
\end{align}
\end{subequations}
corresponding to the closure relations of Eqs.~\eqref{eqn:b_closures}.

We mention that the indirect correlation function,
$\eta(1,2)=h(1,2)-c(1,2)$, that appears in Eqs.~\eqref{eq:op_her},
\eqref{eq:closure} is computed at each iteration step from the OZ
equation, and the expansion \eqref{eqn:O_ex} is performed with the
function $\eta(1,2)$.

The algorithm is written using the angular components of the operator
$\op{O}[\cdot]$, Eq.~\eqref{eq:op_her}, and of the correlation functions
$c(1,2)$, $\eta(1,2)$ as described in full detail in Ref.~\cite{patey1987a}.

In the numerical calculation, the expansion in rotational
invariants of the correlation functions, Eq.~\eqref{eqn:labf},
is truncated at $m_{\text{max}}=n_{\text{max}}=8$ and all
non-zero components consistent with this truncation are kept. The
integral equation was discretized on a grid in steps of $0.01b$.

The first- and second-order bridge diagrams were computed using an
extension of the Monte Carlo methods described by Ree and Hoover
\cite{ree64,ree67}. The first step was to convert the diagrams from
Ref.~\cite{patey1989a}, given in terms of Mayer $f$-bonds, into
Ree-Hoover diagrams, where field points are connected either by an
$f$-bond or by an $e$-bond, where $e=1+f$. The overall bridge function
is obtained from a weighted sum of these Ree-Hoover diagrams. Particle
$1$ is placed at the origin with its symmetry axis along the $z$ axis,
and a second particle is placed at random so that it overlaps the
first particle. A third particle is similarly randomly placed to
overlap the second particle and so on.  When calculating the first set
of bridge diagrams, a chain of four such particles is generated. The
second set of bridge diagrams require a five-particle chain.  The
overlaps between all pairs of particles are checked.  If the
configuration corresponds to one of the Ree-Hoover bridge diagrams,
then the separation between the two end particles of the chain is
calculated, ready for accumulation as a histogram. To obtain the
angular expansion coefficients, the Ree-Hoover weighting is multiplied
by the spherical harmonic product $Y^{*}_{m\chi}(\bm{u}_1)
Y^*_{n\bar{\chi}}(\bm{u}_2)$, where the unit vectors are expressed
relative to the vector joining the two end particles of the chain (1
and 2). The components of the bridge diagrams are thus in the
molecular frame.  After a sufficient number of Monte Carlo
configurations have been generated, $N_\text{conf}$, the final results
for the bridge function are obtained by normalising the histogram
values, firstly by a factor of $N_\text{conf}$, secondly by a factor
of the volume of the spherical shell corresponding to the separation
between the particles and thirdly by an appropriate power of the pair
excluded volume (i.e.\ the square for the first-order term and the
cube for the second-order term). Errors may be estimated in the
standard way, by dividing the total number of configurations into
sub-batches and calculating sub-averages.

We used $1.6\times 10^9$ trial chain configurations to obtain the
first bridge diagram and $1.1\times 10^{9}$ trial chain configurations
to obtain the second bridge diagram. The relative error estimate is
close to $1 \%$ except for the $r<0.1b$ domain that is sampled poorly by
this method.

In summary we now have four sets of integral equation results with which to
compare simulation, corresponding to the four closures of
Eqs.~\eqref{eqn:b_closures}, namely PY, HNC, HNC+B2 (first-order bridge) and
HNC+B3 (first- and second-order bridges).

\section{\label{sec:res}Results}

\subsection{Equation of state and stability with the bridge diagrams}

\begin{figure}
 \includegraphics[width=1.0\linewidth]{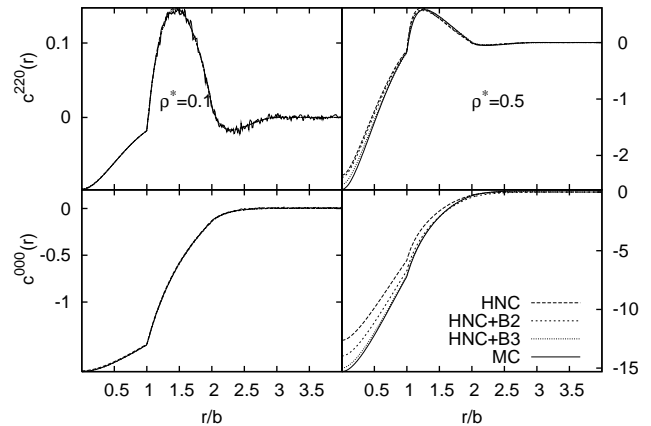}%{c_of_r.eps}
\caption{\label{fig:dcf} The direct correlation function components
$000$ and $220$ (in the lab frame) at reduced densities $\rho^*=0.1$
(left) and $\rho^*=0.5$ (right) obtained from IET and MC.}
\end{figure}

The angular coefficients of the direct and the indirect correlation
functions obtained from PY and HNC integral equations for
non-spherical particles have already been extensively compared with
simulation results in Refs.~\cite{patey1987a,patey1988a, talbot90}. We
limit our discussion to the effect of the inclusion of the bridge
diagrams in the closure.  Fig.~\ref{fig:dcf}, for two angular
components of the direct correlation function at two densities, shows
that the agreement between MC data and IET improves at high density if
the HNC closure is supplemented by the inclusion of the low-order
bridge diagrams.

\begin{figure}
 \includegraphics[width=1.0\linewidth]{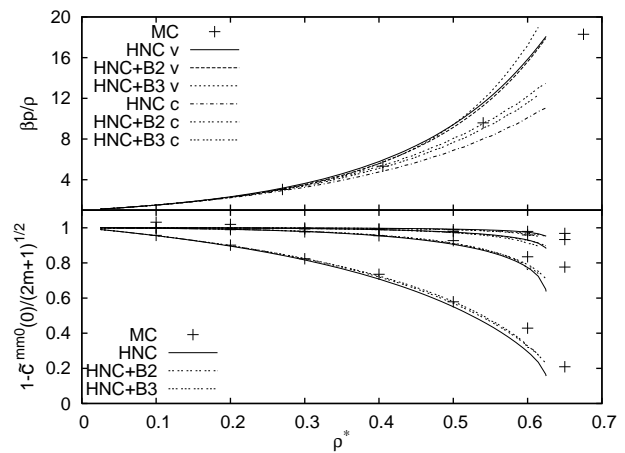}%{eq_of_state.eps}
\caption{\label{fig:eos} The equation of state (upper panel) 
obtained by the virial (v) and compressibility (c) routes; and
the Kerr coefficients (lower panel) obtained from pure HNC
equation and HNC with the first two bridge diagram corrections. The
lines for the Kerr coefficients correspond to $m=2,4,6,8$ in
ascending order and the symbols represent the MC data (equation of
state data from Ref.~\cite{pospisil1993a}.)}
\end{figure}

We have a mixed picture for the equation of state and Kerr or
stability coefficients. The latter gives a measure the stability of the
isotropic phase relative to the nematic phase. The isotropic phase
is stable when \cite{stecki79,stecki81}
\begin{equation}
  1-(2m+1)^{-1/2}\tilde{c}^{mm0}(0) > 0 \;,\quad m=2,4,6,\ldots
\end{equation}
where $\tilde{c}^{mm0}(0)$ is the low-$k$ limit of the Fourier-transformed
 direct correlation function component $c^{mm0}(r)$ in the laboratory frame.
 Fig.~\ref{fig:eos} shows that the inclusion of the first-order bridge
diagram improves the agreement with the MC data for both the virial
and compressibility pressures; the compressibility pressure, in
particular, follows the MC results very closely. Surprisingly, the
inclusion of the second-order bridge diagram increases the deviation
of the pressure from MC at high densities. The same figure shows that
the $m=2$ Kerr coefficient agrees more closely with the simulation
results if bridge corrections are included, but the change is less
clear for $m>2$.

\subsection{\label{ssec:res_cav}Cavity correlation function}

\begin{figure}
 \includegraphics[width=\linewidth]{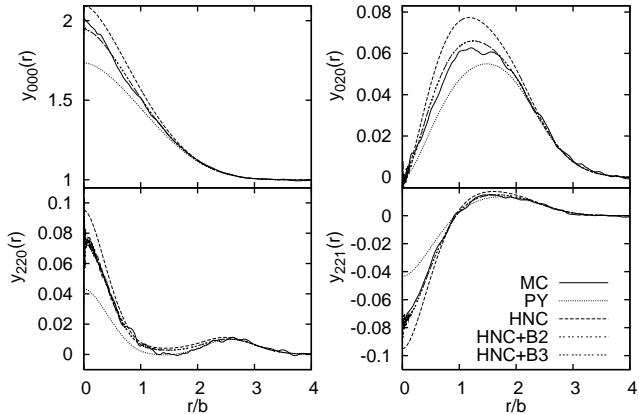}%{y_of_r_rho0.10_LA.eps}
\caption{\label{fig:y_la10} The components of the $y$ function at
density $\rho^*=0.1$  obtained from MC
and IET: PY, HNC, HNC+B2 and HNC+B3.}
\end{figure}

\begin{figure}
  \includegraphics[width=\linewidth]{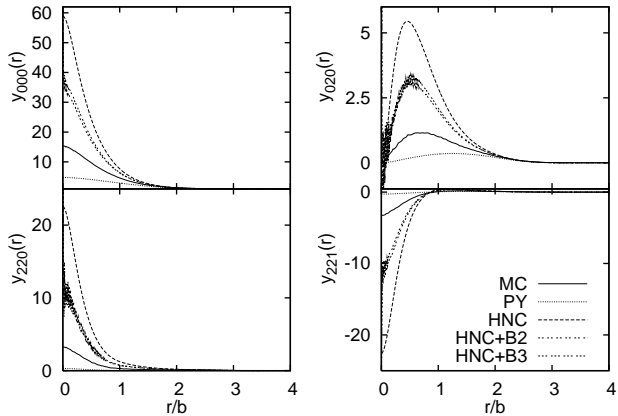}%{y_of_r_rho0.30_LA.eps} 
  \caption{\label{fig:y_la30} The same functions as in
  Fig.~\ref{fig:y_la10} at density $\rho^*=0.3$.}
\end{figure}

Before presenting our numerical results, it is worth considering some
exact, analytical properties of the cavity function at $r=0$.  Firstly
we note that at $r=0$, the cavity function only depends on the
relative orientations of the two particles, 1 and 2, and thus may be
expanded in terms of Legendre functions of $\bm{u}_1 \cdot
\bm{u}_2$. Using the spherical harmonic addition theorem and comparing
the results with Eq.~\eqref{eq:ymnx}, one finds that $y_{mn\chi}(0)$
is zero unless $m = n$. Furthermore $y_{mm\chi}(0) = (-)^\chi
y_{mm0}(0)$. It may be seen from the
figures that our calculated functions obey this condition to within
statistical error. These properties result from the fact that the
cavity function is well-behaved at $r = 0$ and similar conditions
exist for the components of the direct correlation function and bridge
function at $r = 0$.

Secondly we note that it has been shown for hard spheres that the
cavity function at $r = 0$ is related to the excess chemical potential
of the fluid, whilst the gradient of the cavity function at $r = 0$ is
related to the pressure \cite{meeron1968a,henderson1983a}. These
calculations may be generalised for anisotropic hard bodies and we
obtain the exact results (for axially symmetric particles)
\begin{align}
    y(\bm{u},\bm{u},r=0)& = \exp(\beta\mu_{\text{ex}}) \\
    \frac{dy(\bm{u}, \bm{u},r)}{dr}\Big\vert_{r=0} & =
    -\frac{1}{4\pi}\rho y(\bm{u},\bm{u},0) \int_{u_{1,z} \ge
    0} d\bm{u}_1\; u_{1,z} \notag \\ & \int d\bm{u}_2 \;
    r_{c}^2(\bm{u}_1,\bm{u}_2)
    g(r_\text{c},\bm{u}_1,\bm{u}_2) \ , \label{eqn:y_der}
\end{align}
where the integral over $\bm{u_1}$ is restricted to the positive
region of its $z$ component; $r_\text{c}(\bm{u}_1,\bm{u}_2)$ is the
contact distance of the two ellipsoids and
$g(r_\text{c},\bm{u}_1,\bm{u}_2)$ is the contact value of the pair
distribution function at the given orientation.  In the special case
of hard spheres (i.e.\ $r_\text{c}=$ constant), Eq.~\eqref{eqn:y_der}
gives the aforementioned relationship with the pressure, but in
general, so far as we can see, there is no simple connection between
the r.h.s. of Eq.~\eqref{eqn:y_der} and any thermodynamic property of
the fluid.

We now turn to our numerical results. Selected spherical harmonics
components of $y(1,2)$ are shown in
Figs.~\ref{fig:y_la10}-\ref{fig:y_la30} for two densities
$\rho^*=0.1,\; 0.3$. The most obvious conclusion is that both the PY
and HNC predictions differ greatly from the simulation results as the
density increases. In general the PY predictions are far too small in
magnitude, whereas the HNC results are far too big. This is
particularly evident for the isotropic component, $y_{000}(r)$, at low
values of $r$, where $y_{000}(r)$ rises dramatically. At higher
densities HNC and simulation coefficients differ by several orders of
magnitude (from simulation $y_{000}(0)= 2050.1$, while from HNC theory
$y_{000}(0)=3033586.2$ for $\rho^*=0.50$).

The inclusion of the bridge diagrams in the HNC closure improves
significantly the agreement with MC at $\rho=0.1$, see
Fig.~\ref{fig:y_la10}, but theory is still far from simulation for
$\rho=0.3$, see Fig.~\ref{fig:y_la30}.

\subsection{\label{ssec:rec_brid}Bridge function}

\begin{figure}
\includegraphics[width=\linewidth,clip]{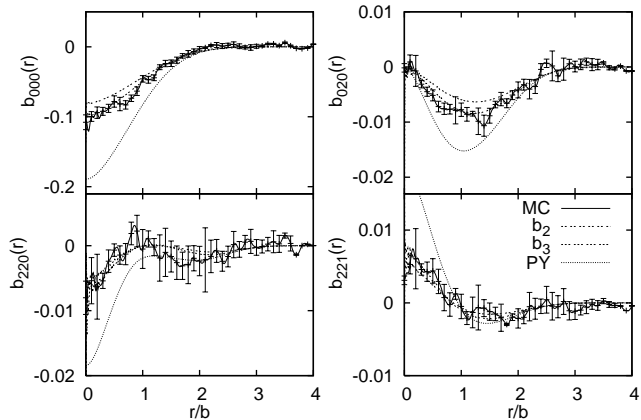}%{b_of_r_rho0.10_LA.eps}
\caption{\label{fig:b_d=10} Spherical harmonics components of $b(1,2)$
for $\rho^*=0.10$ found from simulation, PY theory , and the virial
expansion truncated at 2nd ($b_2$) and 3rd order ($b_3$).}
\end{figure}

\begin{figure}
\includegraphics[width=\linewidth,clip]{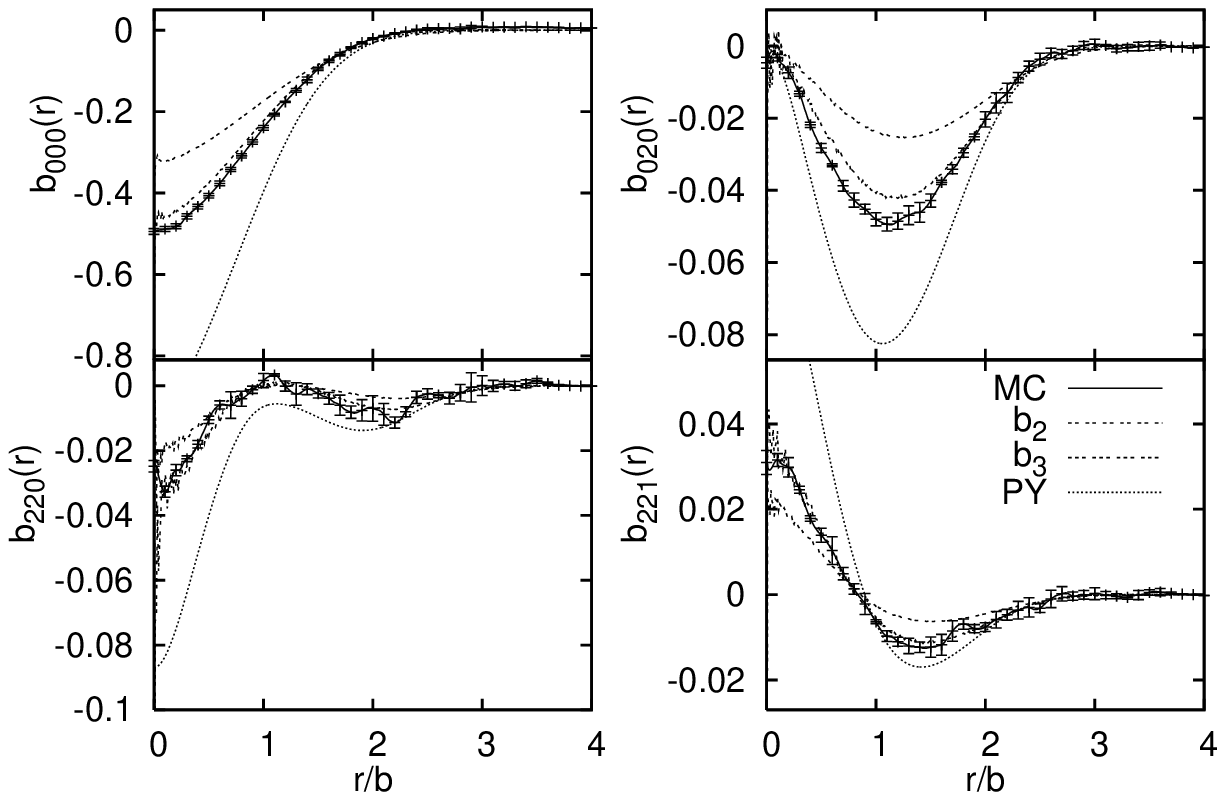}%{b_of_r_rho0.20_LA.eps}
\caption{\label{fig:b_d=20} $\rho^*=0.2$ with same functions as in
Fig.~\ref{fig:b_d=10}.}
\end{figure}

\begin{figure}
  \includegraphics[width=\linewidth,clip]{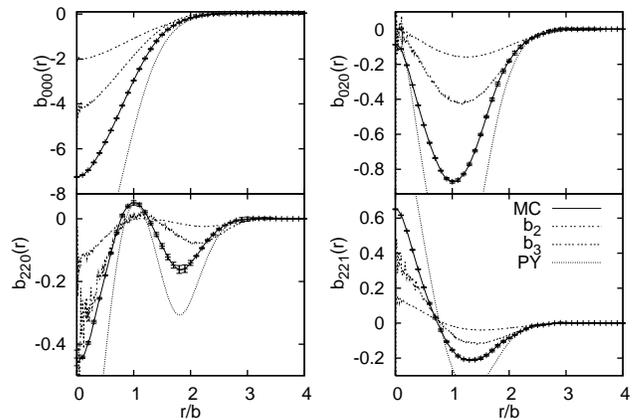}%{b_of_r_rho0.50_LA.eps}
  \caption{\label{fig:b_d=50}$\rho^*=0.5$ with same functions as in
Fig.~\ref{fig:b_d=10}.  }
\end{figure}

Shown in Figs. \ref{fig:b_d=10}-\ref{fig:b_d=50} are selected bridge
function components calculated from simulation, PY and virial
expansion truncated at  the second order ($b_2(1,2)$) and third order
($b_3(1,3)$).

As can be seen the PY $b_{000}(r)$ is always larger than the
simulation $b_{000}(r)$ by approximately a factor of two. This seems to be independent of density.  The shape of
this component, both from simulation and PY theory, is similar to that
of $b(r)$ calculated for simple fluids
\cite{chapman1992a,chapman1994a}. The slope of
$b_{000}(r)$ goes toward 0 as $r$ goes to 0. This is similar to the
behaviour seen for $b(r)$ for Lennard-Jones and soft sphere systems
\cite{chapman1992a,chapman1994a}, while for the HS fluid $b(r)$
approaches $r=0$ almost linearly \cite{henderson1995a}. PY theory
similarly overestimates the angular $b_{mn\chi}(r)$ coefficients.

Shown in Figs.~\ref{fig:b_d=10}-\ref{fig:b_d=50} are also the bridge
function calculated from simulation and from the first and second
terms in the diagrammatic expansion. As can be seen at the lowest
density studied ($\rho^*=0.10$) the first-order expansion gives
reasonable agreement with the simulated bridge function components,
although they are underestimated relative to simulation. Adding the
second term in the expansion improves the agreement quite
considerably. At a higher density, $\rho^*=0.20$, the agreement is
less good, with the first-order expansion seriously underestimating
the coefficients. Again the agreement improves with the addition of
the second term, in fact this diagrammatic approximation of the bridge
function is better overall than the approximation obtained from the PY
equation. As the density increases the agreement between MC results
and the truncated virial expansion worsens, see Fig.~\ref{fig:b_d=50}.

\section{\label{sec:conc}Conclusions}

In this paper we have presented the calculation of the pair
correlation functions $h(1,2)$, $c(1,2)$, $y(1,2)$, and $b(1,2)$ for
the spheroid fluid from both simulation and IET. The total and direct
correlation functions have been calculated using methods previously
described \cite{patey1987a,allen1995a}. The cavity function was
calculated from simulation using a combination of a direct simulation
method and a test-particle approach. In order to improve the sampling
of $y(1,2)$ in the direct simulation approach an umbrella sampling
scheme using a weight function determined iteratively during the
simulation itself is employed.  From IET the cavity function is
determined directly using the approximate closure relations.

Comparison between simulation and integral equation show, as reported
before \cite{patey1987a,patey1988a,talbot90}, reasonable agreement
between the coefficients of the total and direct correlation
functions.  However theory predicts the simulated cavity function
poorly, with PY theory underestimating and HNC theory overestimating
$y(1,2)$ within the overlap region. This error rapidly increases with
density, leading to, at the highest densities studied, errors of
several orders of magnitude.

 The bridge function calculated from the truncated virial expansion is
in good agreement with MC results at low density but significant
differences appear as the density increases.  The bridge function
calculated from PY theory follows the general shape of the MC results
but the quantitative agreement is poor.

To the best of our knowledge this work presents the first calculation
of both the full bridge and cavity functions for molecular fluids from
simulation. As the approximate closure relations used in integral
equation theory correspond to approximations to the bridge function,
knowledge of its exact form will, hopefully, be of great benefit in
developing improved theories of molecular fluids.

\section*{Acknowledgements}

LA and AJM thanks Aurelian Perera for useful discussions on integral
equation numerical techniques. This work was supported by EPSRC grants
GR/S77240 and GR/S77103 Computational resources were provided by the Centre for
Scientific Computing, University of Warwick and Manchester Computing,
University of Manchester.

\appendix*
\section{Wang-Landau sampling}

Consider a system with a property $X$. The probability of finding the
system with a particular $X=X_1$ is given by a probability
distribution $p(X)$. In many cases this distribution is peaked around
certain values of $X$, meaning that in a standard simulation values
away from these are likely to be poorly sampled. When it is desirable
to get information about these unlikely states it is common to apply a
weight function, $g(X)=\exp(-\beta W(X))$, that changes the standard
Metropolis acceptance criteria to
\begin{equation}\label{eqn:wl1}
  p(X_1\rightarrow X_2)=\frac{g(X_1)}{g(X_2)}\exp\left[
  -\beta(E(X_2)-E(X_1))\right].
\end{equation}
The simulated probability distribution $p_\text{sim}(X)$ then becomes
\begin{equation}\label{eqn:wl2}
  p_\text{sim}(X)=p(X)g(X).
\end{equation}
Ideally the effect of the weight function is to make the simulation 
probability distribution flat, i.e. $p_\text{sim}(X)=1$, which implies
\begin{equation}\label{eqn:wl3}
  W(X)=\frac{1}{\beta}\log p(X),
\end{equation}
Of course the $W(X)$ needed to achieve this perfectly flat histogram
is not known in advance, otherwise the probability distribution would 
also be known in advance, thus rendering the actual act of performing
the simulation somewhat redundant.   The problem has then become one
of determining the weight function needed to produce a flat histogram.

At the start of the simulation the weight function is initially set to 
be constant, i.e. $g(X)=1, W(X)=0$. After each attempted MC move 
$X_1\rightarrow X_2$ (made using the modified criteria Eq.~\ref{eqn:wl1}) the weight function for the resulting state 
$X_{1/2}$ (either $X_1$ or $X_2$) is multiplied by a modification factor,
\begin{align*}
  g(X_{1/2})&\rightarrow fg(X_{1/2}) \\
  W(X_{1/2})&\rightarrow W(X_{1/2})+\log f \;.
\end{align*}
Simultaneously the probability histogram $p_\text{sim}(X_{1/2})$ is also
incremented. This continues with the weight function and probability
being updated after every attempted change in $X$ until probability
histogram is flat. The flatness condition may be defined in several
ways and will be discussed momentarily. Once this condition has been
reached the probability histogram is reset to zero and $f$ is
modified. Typically
\begin{align*}
  f&\rightarrow \sqrt{f} \\
  \log f&\rightarrow \frac{1}{2}\log f \;.
\end{align*}
This then continues until the modification factor becomes close to 1 ($\log f$ 
gets close to machine precision). The final $p(X)$ may then be determined from
\begin{equation}
  p(X)=p_\text{sim}(X)/g(X)=p_\text{sim}(X)\exp\left(\beta W(X)\right) \ .
\end{equation}

A few general notes on the method are due. First updating the weight
function during the simulation may be seen to violate the principle of
detailed balance. However, this is most severe at the beginning of the
simulation. As $f$ tends toward 1, the changes in the weight function
become increasingly small. It has been shown that a viable MC scheme
need only asymptotically obey detailed balance
\cite{deem2005a}. Additionally once a sufficiently good weight
function has been determined, the simulation may be continued without
updating the weight function and statistics may be gathered from this
\cite{depablo2002a}.  Secondly as a perfectly flat histogram is
unlikely to be reached during a finite simulation the flatness
condition may be seen to be somewhat arbitrary. In the first
implementations the histogram was declared flat when the smallest
$p_\text{sim}(X)$ was within a given percentage of the average.  However it
is not impossible to imagine pathological distribution (e.g.\  with a
few large narrow peaks) that are far from flat but still fulfil this
criteria. An alternative is to update $f$ whenever every bin has been
visited a minimum number of times. While this may appear less rigorous
then the first method, as the simulation progresses and $W(X)$ becomes
closer to $(1/\beta)\log p(X)$ then $p_\text{sim}(X)$ should become
flat. Additionally this ensures that $p_\text{sim}(X)$ has a chance to
adjust to the new $f$ and avoids any spurious early updates. One final
point is that the Wang-Landau method was originally formulated for
systems with discrete degrees of freedom (specifically the Ising
model). When $X$ is continuous the probability histogram and weight
functions are calculated for bins of finite width $X,\;X+ \delta X$
and bin width may become a perturbing factor in the results.

In the present problem we are interested in the probability distribution of 
a pair of non-interacting particles. The variable of interest is the radial 
separation of these particles $r_{12}$, which is discretized into bins
of width $\delta r=0.01$. As mentioned before the $r_{12}$ 
range is divided into a set of overlapping windows. A Wang-Landau simulation 
is used to determine the weight function to produce a constant $p(r_{12})$ 
within each window. $f$ is updated whenever $p(r_{12})$ fulfils two
criteria: i) the largest difference between any bin and the average is
less than 10\% and ii) the smallest values of any bin is 100.

\end{document}